# Unveiling exotic magnetic phase diagram of a non-Heisenberg quasicrystal approximant

Farid Labib[1*], Kazuhiro Nawa[2*], Shintaro Suzuki[3], Hung-Cheng Wu[2], Asuka Ishikawa[1], Kazuki Inagaki[4], Takenori Fujii[5], Katsuki Kinjo[2], Taku J. Sato[2] and Ryuji Tamura[4]

[1] Research Institute of Science and Technology, Tokyo University of Science, Tokyo 125-8585, Japan

[2] Institute of Multidisciplinary Research for Advanced Materials (IMRAM), Tohoku University, Sendai 980-8577, Japan

[3] Department of Physical Science, Aoyama Gakuin University, Kanagawa 252-5258, Japan

[4] Department of Material Science and Technology, Tokyo University of Science, Tokyo 125-8585, Japan

[5] Cryogenic Research Center, The University of Tokyo, Bunkyo, Tokyo 113-0032, Japan

[*]Corresponding authors:
labib.farid@rs.tus.ac.jp
kazuhiro.nawa.e5@tohoku.ac.jp

**Abstract**

A magnetic phase diagram of the non-Heisenberg Tsai-type 1/1 Au-Ga-Tb approximant crystal (AC) has been established across a wide electron-per-atom ($e/a$) range via magnetization and powder neutron diffraction measurements. The diagram revealed exotic ferromagnetic (FM) and antiferromagnetic (AFM) orders that originate from the unique local spin icosahedron common to icosahedral quasicrystals ($i$QCs) and ACs; The noncoplanar whirling AFM order is stabilized as the ground state at the $e/a$ of 1.72 or less whereas a noncoplanar whirling FM order was found at the larger $e/a$ of 1.80, with magnetic moments tangential to the Tb icosahedron in both cases. Moreover, the FM/AFM phase selection rule was unveiled in terms of the nearest neighbour ($J_1$) and next nearest neighbour ($J_2$) interactions by numerical calculations on a non-Heisenberg single icosahedron. The present findings will pave the way for understanding the intriguing magnetic orders of not only non-Heisenberg FM/AFM ACs but also non-Heisenberg FM/AFM $i$QCs, the latter of which are yet to be discovered.

## 1. Introduction

Quasicrystals, as aperiodically ordered intermetallic compounds, are among the latest discoveries in the field of condensed matter physics which have attained considerable attention due to their exotic physical properties. In particular, Tsai-type icosahedral quasicrystals (*i*QCs) and their cubic approximant crystals (ACs) embrace a new type of strongly correlated electron systems with a number of interesting properties such as recent quasiperiodic long-range magnetic orders [1,2] and an unconventional quantum critical phenomenon [3], to count a few. Their atomic structure commonly comprise a multi-shell polyhedron with the outermost shell being a rhombic triacontahedron (RTH) and the inner shells being an icosidodecahedron, an icosahedron, a dodecahedron, and a central unit, respectively, as schematically shown in Fig. 1a. The vertices of the icosahedron shell are exclusively occupied by rare earth (*RE*) elements.

Recently, there has been growing interest in the magnetic properties of AC phases, not only because of their close association with *i*QCs but also due to the theoretical discovery of exotic magnetic orders such as hedgehog, anti-hedgehog, whirling, and anti-whirling, etc, with a variety of topological numbers [4]. The lowest order AC, i.e., 1/1 AC, typically grows in the space group $Im\bar{3}$ [5], where the RTH clusters are arranged in the form of body-centered cubic structure. Figure 1b displays the configuration of *RE* elements (all symmetrically equivalent) within one unit cell of 1/1 AC [6]. The interplay between the spins and the itinerant electrons often leads to spin-glass-like freezing behavior [7,8] due to the spin competition inherent to such a complex atomic environment even though long-range ferromagnetic (FM) [1,2,9–14] and antiferromagnetic (AFM) [15–21] orders do establish under circumstances, some of which are still unknown.

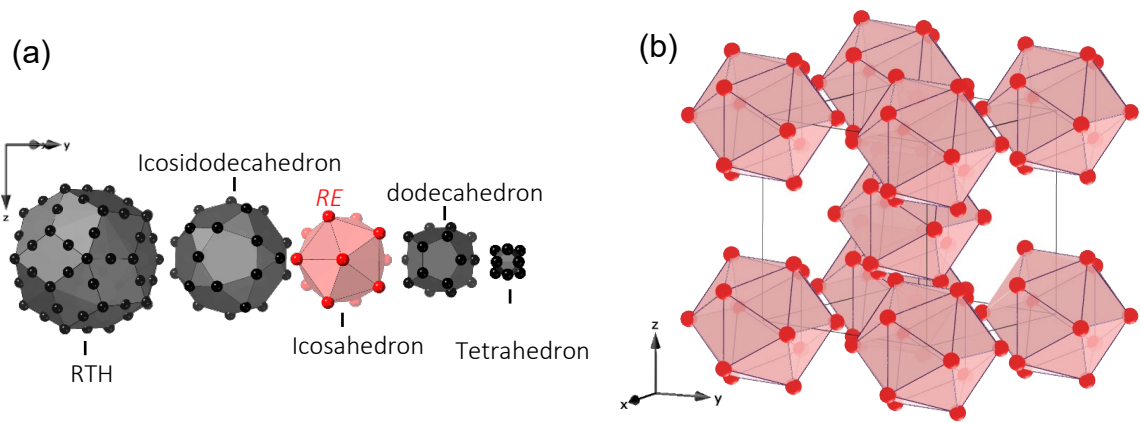

FIG. 1. (a) A typical shell structure of the Tsai-type icosahedral quasicrystal (*i*QC). From the outermost shell to the center: a rhombic triacontahedron (RTH) with 92 atoms sites, an icosidodecahedron (30 atomic sites), an icosahedron (12 atomic sites), a dodecahedron (20 atomic sites) and inner tetrahedron (4 atomic sites). (b) Typical arrangement of *RE* sites within the unit cell of 1/1 AC.

As one of the major accomplishments in this area, electron per atom (*e/a*) dependency of the paramagnetic Currie-Weiss ($\theta_w$) temperature in the Au-based Tsai-type 1/1 ACs has been verified [22,23], as schematically shown in Fig. 2. In the figure, where $\theta_w$ has been normalized by *dG*, i.e., de Gennes parameter expressed as $(g_J-1)^2 J(J+1)$ where $g_J$ and $J$ denote the Landé g-factor and the total angular momentum, an oscillatory behavior of the $\theta_w$ with $e/a$ is evident indicating dependency of the magnetic ground states to the Fermi energy in ACs. Notice that the boundaries between the magnetic ground states are subject to slight shift with $e/a$ depending on the constituent elements of the alloy system resulting in relatively wide inter-state boundaries shown in Fig. 2. As far as the magnetic structure of the ordered states is concerned, for the AFM state, recent neutron diffraction experiments on the $Au_{72}Al_{14}Tb_{14}$ [18], $Au_{65}Ga_{21}Tb_{14}$ 1/1 ACs [24] and $Au_{70}Al_{16}Tb_{14}$ [25], suggest noncoplanar spin configuration on the icosahedral clusters whirling around a crystallographic [111] axis. Likewise, for the FM state, neutron diffraction experiment on the $Au_{70}Si_{17}Tb_{13}$ [26] and Au-Si-RE (RE = Tb, Ho) [27] 1/1 ACs revealed noncoplanar FM spin configuration on the icosahedron clusters [18,26].

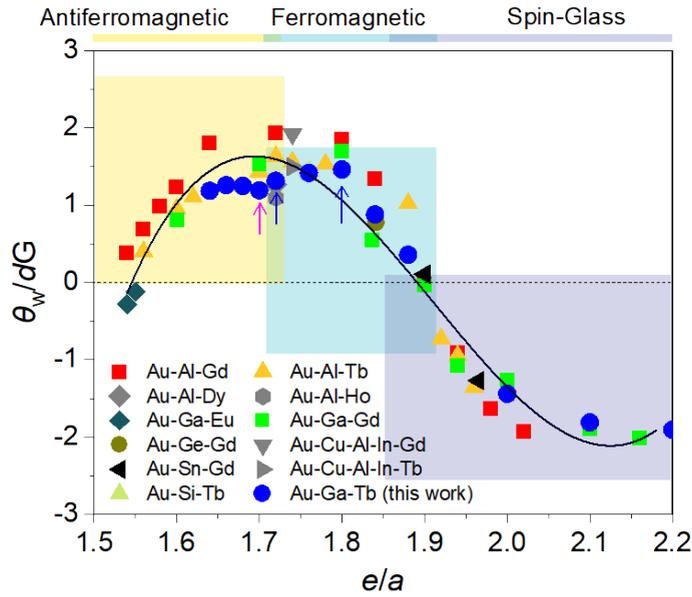

FIG. 2. Variation of normalized Weiss temperature $\Theta_w/dG$ and the magnetic ground states versus the electron-per-atom (*e/a*) ratio spanning from 1.5 and 2.2. The oscillatory behavior of the polynomial fitting to the $\Theta_w/dG$ has been observed in the Au-based ACs. The blue filled circles correspond to the data points derived from the present study. The blue arrows mark the samples that are further examined by neutron diffraction experiment in the present work while the purple arrow indicate the sample that was examined by neutron diffraction experiment in ref. [24].

The present study is set to determine $e/a$ dependence of the magnetic structures in the non-Heisenberg Tsai-type 1/1 Au-Ga-Tb AC through magnetization measurements over a broad range of $e/a$ ratios ranging from 1.6 to 2.2 and powder neutron diffraction (PND) experiments at particular $e/a$ values corresponding to the single FM regime and the FM/AFM border, respectively. The PND results suggested a single noncoplanar FM order at $e/a = 1.80$ with out-of-mirror plane magnetic component and canting angle of 50(5) degrees, while at $e/a = 1.72$ successive FM and AFM transitions at $T = 11.2$ and 9.9 K, respectively, are noticed confirming the existence of an intermediate phase. Based on the findings, an exotic magnetic phase diagram of the non-Heisenberg Tsai-type 1/1 AC has been developed for the first time.

## 2. Material and methods

Polycrystalline 1/1 AC alloys with nominal compositions of $Au_{68-x}Ga_{18+x}Tb_{14}$ ($x = 0-27$) were prepared employing the arc-melting technique followed by isothermal annealing at $T = 973$ K for 100 hours. The bulk magnetization of selected samples was examined and two of them with nominal compositions of $Au_{60}Ga_{26}Tb_{14}$ and $Au_{64}Ga_{22}Tb_{14}$ were selected for further analysis via PND experiments. This selection was based on their critical *e/a* values that locate them inside a single FM and at the FM/AFM border, respectively, as marked by blue arrows in Fig. 2. Powder X-Ray diffraction (XRD) were carried out for phase identification using Rigaku SmartLab SE X-ray diffractometer with Cu-Kα radiation. The dc magnetic susceptibility of the samples was measured under zero-field-cooled (ZFC) and field-cooled (FC) conditions using superconducting quantum interference device magnetometer, Quantum Design: MPMS3 (or PPMS; equipped with a vibrating sample magnetometer), within $1.8 < T < 300$ K and in external dc fields up to $7 \times 10^4$ Oe. The PND experiment was performed using a triple-axis neutron spectrometer 4G-GPTAS installed at the JRR-3 reactor, Tokai, Japan. For that, 2.0 grams of the $Au_{64}Ga_{22}Tb_{14}$ and 3.8 grams of $Au_{60}Ga_{26}Tb_{14}$ 1/1 ACs were packed in the Aluminum foil and shaped into a thin-walled cylinder with the diameter of 10 mm and 11 mm, respectively. The samples were then sealed in the Aluminum can with He exchange gas. They were cooled down to 2.6 K using a closed cycle 4He refrigerator. Neutrons with the wavelength of λ = 2.352 and 2.359 Å were selected for the NPD measurements of $Au_{64}Ga_{22}Tb_{14}$ and $Au_{60}Ga_{26}Tb_{14}$ using pyrolytic graphite (PG) 002 reflections, respectively. A PG filter was installed in the upstream of monochromator to remove higher harmonic neutrons. The horizontal collimations of 40´-40´-40´-open with the double-axis mode were employed to collect intensities.

## 3. Results and discussion
### *3.1. Sample characterization*

The powder XRD patterns of the prepared $Au_{68-x}Ga_{18+x}Tb_{14}$ 1/1 ACs with *x* being in a range of 0 – 27 is provided in Fig. S1 of the Supplementary Information. All the peaks are assigned to the Tsai-type 1/1 AC indicating widely spanned single-phase 1/1 AC phase in the Au-Ga-Tb phase diagram, which is one of the main advantages of the present Au-Ga-Tb system as it allows us to investigate the *e/a* dependence of their magnetism. In Fig. 3, powder XRD patterns and corresponding Le Bail fittings of the two selected 1/1 ACs with nominal compositions of (a) $Au_{60}Ga_{26}Tb_{14}$ and (b) $Au_{64}Ga_{22}Tb_{14}$ are provided. The fitting curves in Fig. 3 are obtained by assuming the space group $Im\bar{3}$ using the Jana 2006 software suite [28]. The measured ($I_{obs}$) and calculated ($I_{cal}$) peak intensities are represented by red and black lines, respectively. The lattice parameters are determined as 14.68(3) and 14.71(2) Å, the latter corresponding to the sample with 64 at.% Au. The excellent agreement between the positions and intensities of the calculated and experimental peaks indicates a high purity of the synthesized 1/1 ACs which confirms their suitability for further magnetization and PND measurements. The following sub-sections will extensively cover the magnetic properties of the $Au_{60}Ga_{26}Tb_{14}$ and $Au_{64}Ga_{22}Tb_{14}$ 1/1 ACs due to their utmost importance (considering their typical $e/a$) to the main purpose of the present work being the establishment of magnetic phase diagram in non-Heisenberg Tsai-type 1/1 ACs. Details about the magnetization of the remaining samples will be provided in the Supplementary Information.

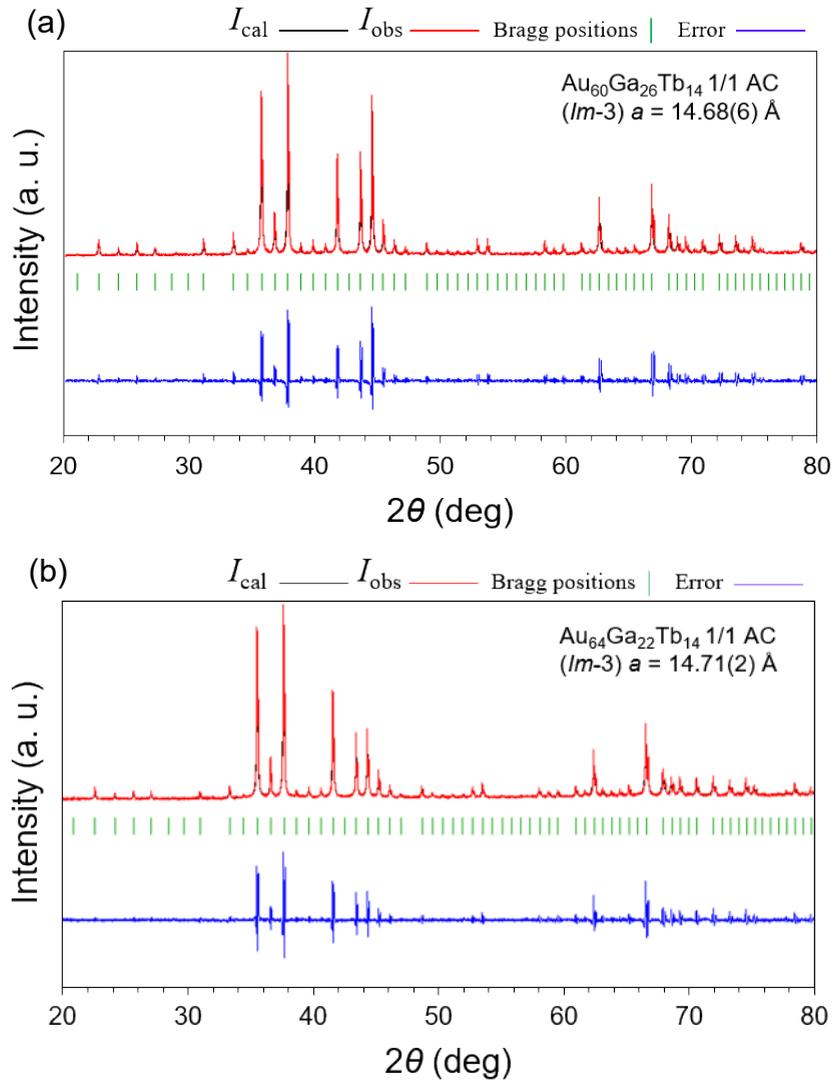

FIG. 3. Le Bail fitting of the powder x-ray diffraction (XRD) patterns of (a) $Au_{60}Ga_{26}Tb_{14}$ and (b) $Au_{64}Ga_{22}Tb_{14}$ 1/1 ACs. The measured ($I_{obs}$), calculated ($I_{cal}$) peak intensities, the difference between the two and the expected Bragg peak positions are represented by red, black, blue lines and green bars, respectively. Overall, excellent agreement between the positions and intensities of the calculated and experimental peaks can be observed indicating high purity of the synthesized samples.

### 3.2. Magnetic properties

*3.2.1. $Au_{60}Ga_{26}Tb_{14}$ 1/1 AC (e/a = 1.80)*

Figure 4a displays temperature dependence of the dc magnetic susceptibility ($M/H$) of the $Au_{60}Ga_{26}Tb_{14}$ 1/1 AC within a temperature range of $T = 0 - 30$ K under FC and ZFC modes, represented by filled and unfilled circles, respectively. In the inset, the high-temperature inverse magnetic susceptibility (defined as $H/M$) of the same compound within a temperature range of 1.8–300 K is provided. The inverse magnetic susceptibility shows a linear behavior well fitted to the Curie–Weiss law:

$$\chi(T) = \frac{N_A \mu_{eff}^2 \mu_B^2}{3k_B(T-\theta_w)} + \chi_0 \tag{1}$$

where $N_A$, $\mu_{eff}$, $\mu_B$, $k_B$, $\theta_w$, and $\chi_0$ denote the Avogadro number, effective magnetic moment, Bohr magneton, Boltzmann constant, Curie-Weiss temperature, and the temperature-independent magnetic susceptibility, respectively. The estimated $\theta_w$ from the extrapolation of a linear least-squares fitting in the temperature range of 100 K < $T$ < 300 K is +13.8 ± 0.6 K with $\chi_0$ approximating zero. The obtained $\mu_{eff}$ is 9.61 $\mu_B$ close to the calculated value for free $Tb^{3+}$ defined as $g_J(J(J+1))^{0.5}$ $\mu_B$ [29] suggesting localization of the magnetic moments on $Tb^{3+}$ ions.

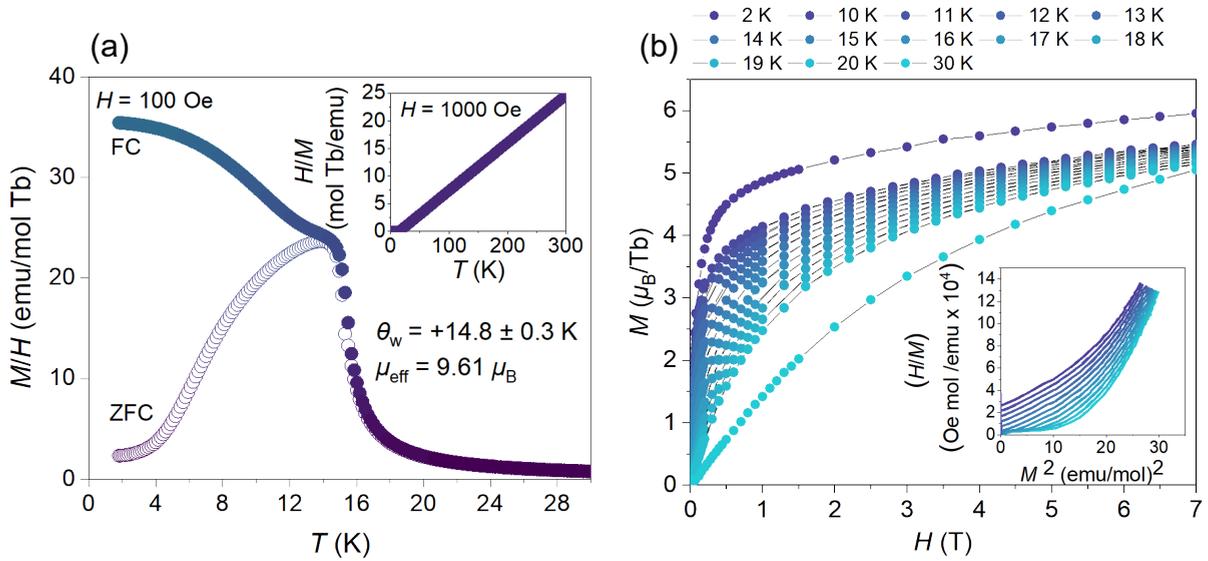

FIG. 4. (a) Low temperature magnetic susceptibility (*M/H*) of the $Au_{60}Ga_{26}Tb_{14}$ 1/1 AC as a function of temperature under FC and ZFC modes and *H* = 100 Oe. The inset shows the inverse magnetic susceptibility (*H/M*) of the same compound. (b) Series of magnetic field (*H*) dependence magnetizations measured within a temperature range of 2 – 30 K up to *H* = 7 T. In the inset of Fig. 4b, Arrott plot (*H/M* vs. $M^2$) of the same sample is provided.

At low temperatures, as seen in Fig. 4a, the magnetic susceptibility rises rather sharply below *Tc* = 15.5 K (estimated from the minimum of the *d*(M/H)/*d*T), which is associated with the pinning of magnetic domain walls during the FC process, a common feature in ferromagnets [9,10]. A series of isothermal field-dependence magnetization (*M-H*) curves of the same compound are recorded in a temperature range of *T* = 2 – 30 K as shown in Fig. 4b. Clearly, at temperatures below *Tc* = 15.5 K, the magnetization increases relatively fast with the applied field being suppressed at ~ 6 $\mu_B$/$Tb^{3+}$ (~ 75 % of the full moment of a free $Tb^{3+}$ ion) at *T* = 2 K and *H* = 7 T. By increasing the measurement temperature, the curvature of the *M-H* curves becomes milder and more like a paramagnetic type. From the Arrott plot curves (i.e., *H/M* vs. $M^2$) of the same sample shown in the inset of Fig. 4b, no negative slope or inflection point can be observed around *Tc* indicating the occurrence of a second-ordered-like transitions in the $Au_{60}Ga_{26}Tb_{14}$ 1/1 AC according to the Banerjee's criterion [30].

Figure 5 presents the PND patterns of the $Au_{60}Ga_{26}Tb_{14}$ 1/1 AC at $T = 2.6$ K and 18 K, i.e., below and above $Tc = 15.5$ K, respectively, wherein the blue vertical bars correspond to the calculated nuclear reflections from the magnetic structure of the $Au_{70}Si_{17}Tb_{13}$ 1/1 AC provided elsewhere [26]. The PND pattern at 18 K is well reproduced by the atomic positions of $Au_{65}Ga_{16}Tb_{14}$ [24] (see Fig. S2 in the supplementary information for details). At the base temperature of $T = 2.6$ K, all the magnetic reflections appear at the allowed angles for the BCC lattice, i.e., $hkl$; $h + k + l = 2n$ ($n$: an integer) with the strongest one being 013 at $2\theta = 29.5°$. The appearance of the strong reflection at $2\theta = 69.7°$ (marked by red arrowhead in Fig. 5) corresponds to the 111 reflection of the aluminium foil used to wrap the powdered sample. Clearly, the PND pattern of the present $Au_{60}Ga_{26}Tb_{14}$ 1/1 AC differs from that of $Au_{70}Si_{17}Tb_{13}$ 1/1 AC [26] in the intensities of the magnetic reflections, which suggests that the orientation of the magnetic moments is different in the two compounds. The intensity of the 031 reflection as a function of the temperature is displayed in Fig. 6. The curve is fitted to a piecewise power law function described as $y = A_1 \times (x-T_c/T_c)^{2\beta} + A_2$ ($x < T_c$); $y = A_2$ ($x > T_c$) in the temperature range of 13–18

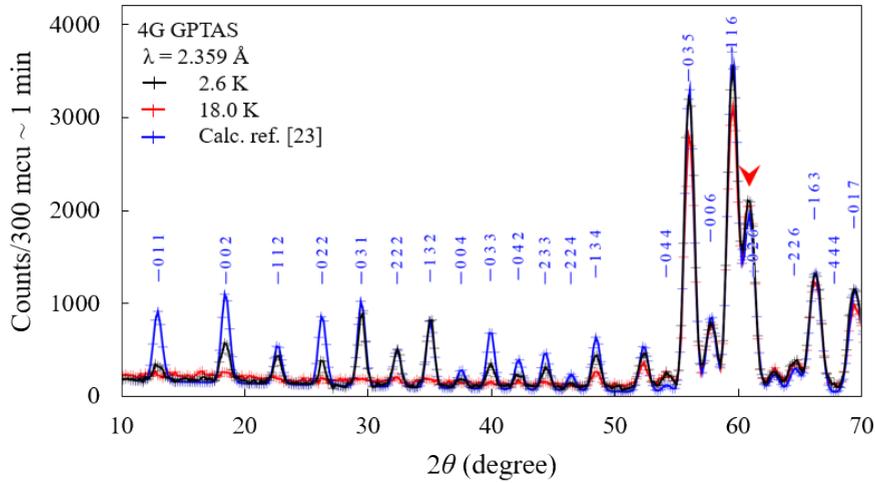

FIG. 5. The powder neutron diffraction (PND) patterns of the $Au_{60}Ga_{26}Tb_{14}$ 1/1 AC at $T = $ 2.6 and 18 K, below and above $T_c = 15.5$ K, respectively, together with the calculated pattern from $Au_{70}Si_{17}Tb_{13}$ 1/1 AC [26]. The blue vertical bars correspond to the calculated nuclear reflections from the magnetic structure of the $Au_{70}Si_{17}Tb_{13}$ 1/1 AC. The arrow at $2\theta = 69.7°$ indicates the peak from the aluminium foil, which is also included in calculation.

K. At high temperatures, the intensity does not change with temperature, but it increases rapidly below $T_c$ = 15.4(2) K, estimated from the above fitting. The critical exponents $\beta$ is estimated to be 0.42(6) from the fitting being in good agreement with $\beta$ = 0.44 in the AFM $Au_{70}Al_{16}Tb_{14}$ [25] and 0.56 in the AFM $Au_{65}Ga_{21}Tb_{14}$ [24]. In addition, in the earlier report in the FM Au-Si-Gd 1/1 AC [31], $\beta$ = 0.47 and $\gamma$ = 1.12 are derived through bulk magnetization measurements and calculations based on the scaling hypothesis [32]. Therefore, the mean-field universality class ($\beta$ = 0.5, $\gamma$ = 1) [33] may hold for the magnetic transitions in a wide variety of 1/1 Acs studied to date.

Next, the magnetic structure analysis was performed using Rietveld refinement under certain restraints. Initial magnetic structures were obtained using magnetic representation theory [34], which assumes that the magnetic structure can be represented using a linear combination of magnetic basis vectors that are part of a single irreducible representation (IR) of the 'k-group' associated with the crystallographic space group. The magnetic modulation vector (0, 0, 0) must be conserved because the magnetic unit cell is identical to the chemical unit cell, which preserves bcc centering-translational symmetry. The magnetic representations of the Tb moments can be simplified into six one-dimensional and two three-dimensional IRs. The magnetic basis vectors (BVs) for all the IRs are obtained as those listed in ref. [26]. Only one three-dimensional irreducible representation (IR7 in in ref. [26]) permits the emergence of spontaneous magnetic moments. However, without applying any restraints, the unique magnetic structure cannot be found because the magnetic reflections induced by the FM order are averaged out in the powder diffraction patterns of the sample with a cubic crystal structure. To deduce a possible magnetic structure, it was assumed that the FM phase belongs to the same magnetic group as other ferromagnetic ACs, i.e., $R\bar{3}$ [27], which has an inherent 3-fold symmetry. The space group of $R\bar{3}$ induces two inequivalent magnetic atoms leading to 6 degrees of freedom for spin arrangement. Additionally, we assumed that the magnetic moment size is the same for Tb sites, which further reduces the degrees of freedom to 3 from IR7. The good agreement was achieved by the Rietveld refinement with 3 adjustable parameters, which is shown by the red solid curves in Fig. 7. The magnetic moment size is estimated as 5.63(22) $\mu_B$, and the refined magnetic structure (as shown in the inset of Fig. 7) has a large component perpendicular to the local mirror plane present in $Im\bar{3}$. The canting angle from the mirror plane is 50(5) degrees, which is much larger than nearly 7 degrees in $Au_{70}Si_{17}Tb_{13}$ 1/1 AC [26]. Moreover, the orientations of magnetic moments are 87(2) degrees canted away from the pseudo 5f-axis.

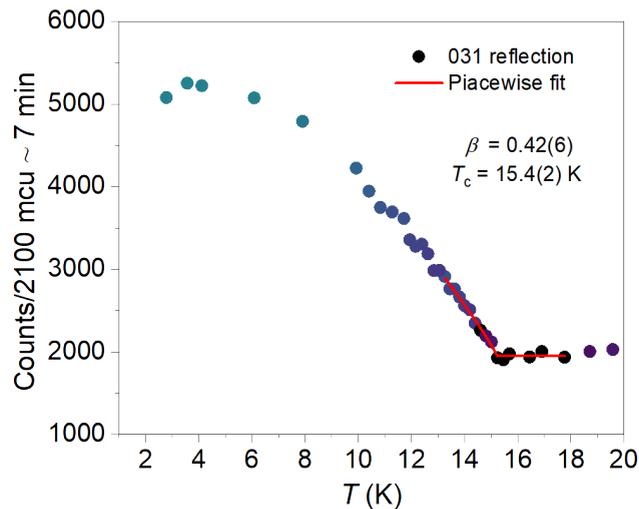

FIG. 6. Temperature dependence of the integrated intensity of the 031 Bragg reflection associated with FM order.

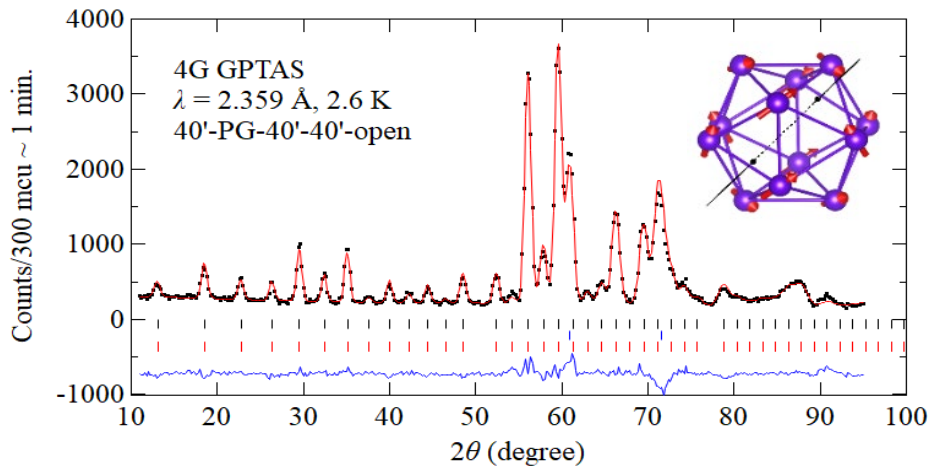

FIG. 7. Rietveld refinement of $Au_{60}Ga_{26}Tb_{14}$ 1/1 AC under some restraints (see text for details). Observed intensities, calculated intensities, and their difference are represented by black dots, red and blue curves, respectively. The position of the nuclear reflections, Aluminum (from sample holder), and magnetic reflections are indicated by black, blue, and red vertical solid lines, respectively. Inset shows the magnetic structure obtained from the refinement.

The observed difference in the magnetic structure of the present compound and that of the $Au_{70}Si_{17}Tb_{13}$ 1/1 AC [26] could be attributed to the difference in their magnetic anisotropy and/or difference in their chemical composition. The CEF calculations using the point charge [4] suggest that ligand atoms' valences affect easy axis magnetization in Tb-based iQCs and ACs, therefore possible chemical disorder induced by compositional difference may stabilize magnetic moments out of the mirror plane by breaking mirror symmetry. Moreover, the higher Tb content (~14 at.%) in the present compound compared to $Au_{70}Si_{17}Tb_{13}$ 1/1 AC [26] could explain some differences in their atomic structure, particularly around the center of the RTH clusters. In the latter, the cluster center is empty [26], while in the present FM $Au_{60}Ga_{26}Tb_{14}$ 1/1 AC, the existence of Tb at the cluster center is quite likely (as is the case in $Au_{65}Ga_{21}Tb_{14}$ [24] and $Au_{70}Si_{16}Tb_{14}$ [12] 1/1 ACs). This could create competing interactions between magnetic moments or increase randomness between FM and AFM interactions through RKKY indirect coupling.

### 3.2.2. $Au_{64}Ga_{22}Tb_{14}$ 1/1 AC (e/a = 1.72)

In this sub-section, magnetic properties of the $Au_{64}Ga_{22}Tb_{14}$ 1/1 AC, the $e/a$ of which equals 1.72 and locates exactly at the boundary between AFM and FM regions (based on the rigid bond approximation shown in Fig. 2) is discussed. The sample is a single phase as evidenced by the powder XRD pattern in Fig. 3b. Figure 8a plots temperature dependence of the magnetic susceptibility of this sample where its rapid rise and subsequent drop can be clearly observed in the course of temperature reduction exhibiting characteristics of both FM- and AFM-like transitions. At paramagnetic region, its inverse magnetic susceptibility shown in the inset of Fig. 8a exhibits a linear behavior well fitted to the Curie-Weiss law in eqn. (1) resulting in $\theta_w$ and $\mu_{eff}$ of +13.8±0.6 K and 9.95 $\mu_B$, respectively. Figure 8b provides temperature dependence of the specific heat divided by temperature ($C_p/T$) where two anomalies at $T_N = 9.9$ K and $T_c = 11.2$ K are noticed. The positions of these anomalies are consistent with the extremums of the *d*(M/H)/*d*T.

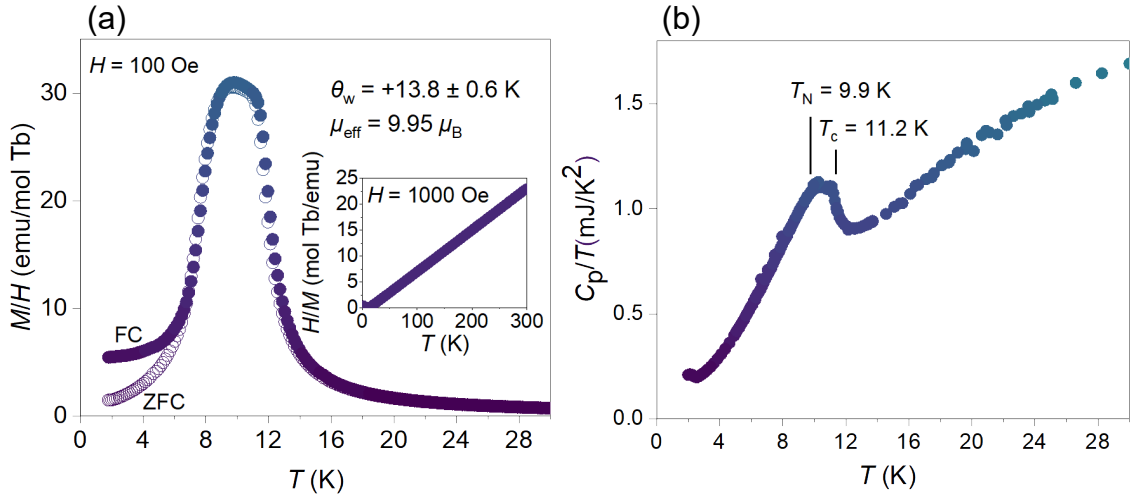

FIG. 8. (a) Low temperature magnetic susceptibility *M/H* of $Au_{64}Ga_{22}Tb_{14}$ 1/1 AC as a function of temperature under FC and ZFC modes. The inset shows inverse magnetic susceptibility *H/M* of the same compound. (b) Normalized temperature dependence of the specific heat capacity within a range of $T = 0 - 30$ K.

In Fig. 9, a field dependence magnetization (*M-H*) of the $Au_{64}Ga_{22}Tb_{14}$ 1/1 AC up to 7 T at 1.8 K (below $T_N$), and 10 K (above $T_N$ and below $T_c$) is shown. At 1.8 K, metamagnetic-like anomalies at $H = 0.21$ T and $H = 0.12$ T (evidenced by the appearance of two critical points in the first derivative of the *M-H* curve in the bottom-right inset of Fig. 9) corresponding to a spin-flop phenomenon during the application and removal of the magnetic field, respectively, are observed. A hysteresis shown in the magnified *M-H* curve in the middle inset of Fig. 9 suggests the spin-reorientation transition to be of first order type [24]. Such hysteresis is also noticed between the FC and ZFC susceptibilities in Fig. 8a, a relatively common feature in a number of AFM 1/1 ACs such as $Cd_6Tb$ [15], and $Au_{65}Ga_{21}Tb_{14}$ [24]. The origin of the hysteresis can be correlated to a spontaneous magnetic moment, the magnitude of which amounts to ~ 0.1 $\mu_B$ (see the bottom-

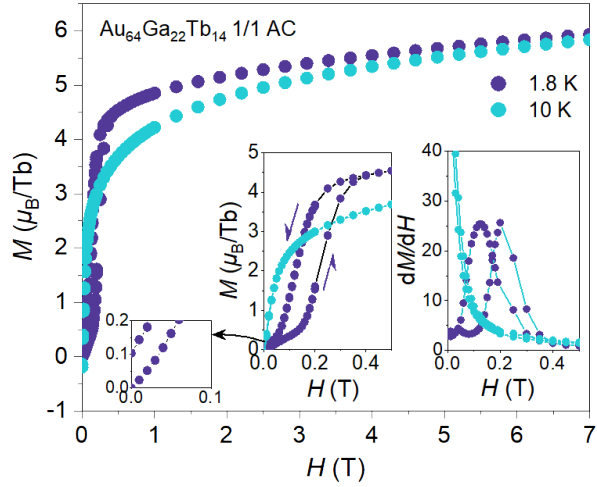

FIG. 9. Field-dependence of the magnetization (*M–H)* of the Au$_{64}$Ga$_{22}$Tb$_{14}$ 1/1 AC at 1.8 and 10 K up to 7 T. The bottom-right inset shows the first derivative of the *M-H* curve. The middle inset represents a magnified view of the *M-H* curves. The far-left inset shows magnified view if the *M-H* curve wherein the spontaneous magnetization is more visible.

left inset of Fig. 9). In addition, the maximum magnetization at $T$ = 1.8 K and $H$ = 7 T approaches to ~ 6 $\mu_B$/Tb$^{3+}$, ~ 75% of the full moment of a Tb$^{3+}$ free ion based on Hund's rule (i.e., 9.00 $\mu_B$/Tb$^{3+}$) most likely due to the existence of uniaxial anisotropy in the Tb$^{3+}$ spins, as suggested by the recent inelastic neutron experiments in the FM Au$_{70}$Si$_{17}$Tb$_{13}$ 1/1 AC [26]. At 10 K, however, while the magnetization reaches ~ 6 $\mu_B$/Tb$^3$, no sign of meta-magnetic anomaly can be detected in its field dependency. These results together with the magnetic susceptibility and specific heat variation suggest the appearance of successive magnetic transitions of AFM- and FM- types at $T_N$ = 9.9 K, and $T_c$ = 11.2 K, respectively

Figure 10 compares PND patterns of the Au$_{64}$Ga$_{22}$Tb$_{14}$ 1/1 AC at $T$ = 2.7, 10 and 15 K. Indices are also shown in the figure. The powder diffraction pattern at 15 K, where magnetic Bragg reflections are absent, is consistent with the atomic positions of Au$_{65}$Ga$_{16}$Tb$_{14}$ [24] (see Fig. S3(b) in the supplementary information for details). On the other hand, at $T$ = 2.7 K they appear at *hkl*;

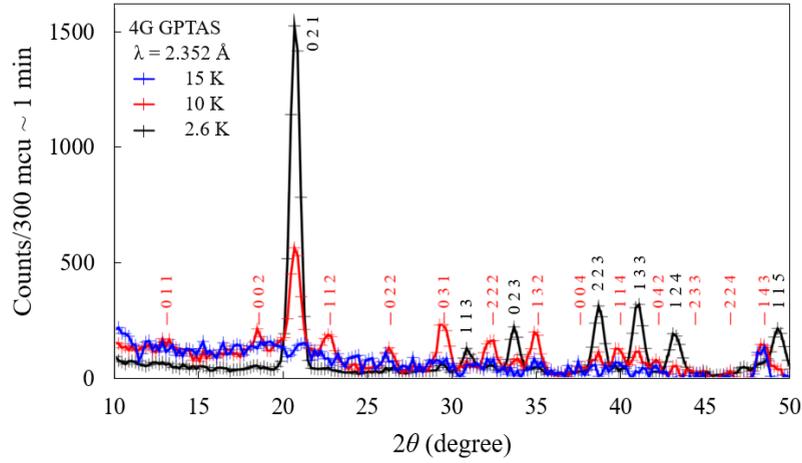

FIG. 10. Low-temperature powder neutron diffraction (PND) patterns of the Au$_{64}$Ga$_{22}$Tb$_{14}$ 1/1 AC at $T$ = 2.7 K (×1/3), 10 K and 15 K. Red vertical bars correspond to the calculated positions of nuclear reflections.

$h + k + l = 2n + 1$ ($n$: an integer) with the strongest one being 021 at $2\theta = 21°$ indicating breaking of the BCC symmetry. A complete consistency can be noticed between the intensity distribution of the reflections observed at $T$ = 2.7 K and those of the AFM Au$_{72}$Al$_{14}$Tb$_{14}$ [18] and the AFM Au$_{65}$Ga$_{21}$Tb$_{14}$ 1/1 ACs [24]. The Rietveld refinement results (shown in Fig. S3(a) and summarized in Table SI in the supplementary information) indicate that the magnetic structure of the AFM phase in the Au$_{64}$Ga$_{22}$Tb$_{14}$ 1/1 AC is essentially isostructural to those reported in Refs. [18,24]. In the AFM order, magnetic moments are required to remain within the mirror plane, which is in contrast with that observed in the FM Au$_{60}$Ga$_{26}$Tb$_{14}$ 1/1 AC in Fig. 7 where the canting angle of 50(5) degrees is noticed between the moments and the mirror plane. The temperature dependence of the magnetic Bragg reflections 021 and 031 are provided in Fig. 11, wherein a rapid rise in the intensity of the 021 reflection below $T_N$ is evident.

In the PND pattern of the Au$_{64}$Ga$_{22}$Tb$_{14}$ 1/1 AC at $T$ = 10 K in Fig. 10, most of the AFM reflections are strongly suppressed. Yet, new Bragg reflections emerge at $hkl$; $h + k + l = 2n$ ($n$: integer) positions, all being crystallography allowed for the BCC lattice, indicating emergence of a FM order in the same sample with just rising the temperature. The strongest magnetic reflection amongst the newly emerged peaks at $T = 10$ K is 031. Note that the intensity distribution of the emerged reflections at 10 K in Fig. 10 is consistent with those of the $Au_{60}Ga_{26}Tb_{14}$ 1/1 AC with a single FM transition at $T = 2.6$ K (see Fig. 5) indicating that the magnetic structure of the FM phases in both compounds (i.e., $Au_{60}Ga_{26}Tb_{14}$ at $T = 2.6$ K ($e/a = 1.80$) and $Au_{64}Ga_{22}Tb_{14}$ at $T = 10$ K ($e/a = 1.72$)) is essentially same. This is indeed confirmed by the Rietveld refinement, whose result is shown in Fig. 12. The refinement was performed by assuming that both the AFM phase found at 2.7 K and the FM phase found for $Au_{60}Ga_{26}Tb_{14}$ 1/1 AC contribute to the powder diffraction pattern. The moment orientation was fixed and only the moment size multiplied by the volume fraction of each phase was adjusted. The two adjustable parameters well reproduce the complicated distribution of the magnetic reflections, supporting the coexistence of the two magnetic phases at 10 K.

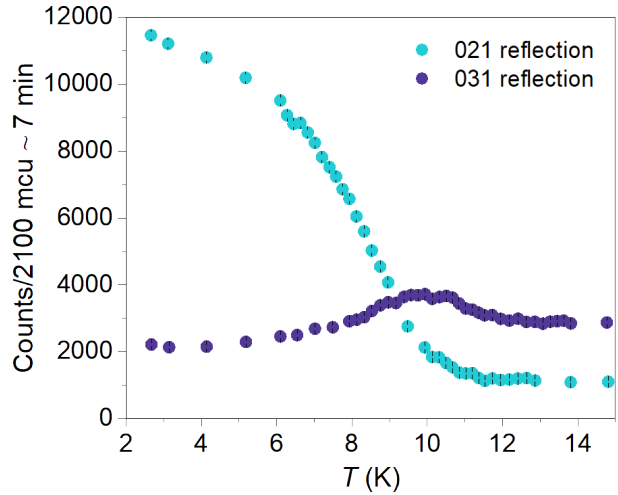

FIG. 11. Temperature dependence of the integrated intensity of the 021 and 031 Bragg reflections.

The coexistence of the magnetic Bragg reflections associated with AFM and FM orders at $T = 10$ K in a single 1/1 AC sample with $e/a$ of 1.72 is a significant outcome that has never been observed in Tsai-type compounds before. This result provides important information about the magnetic phase diagram in Tsai-type non-Heisenberg quasicrystal ACs, as will be discussed later. To our surprise, the intensity of the 031 magnetic reflection (as the strongest one associated with FM order) in Fig. 11 shows a maximum at ~ 10 K and disappears at lower temperatures which implies that the FM phase in the $Au_{64}Ga_{22}Tb_{14}$ 1/1 AC cannot be attributed to the inhomogeneity of the polycrystalline sample and should be treated as inherent characteristic of the emerged FM phase at $e/a = 1.80$. We speculate that the phase coexistence at this particular $e/a$ is due to the competition between intra-cluster nearest neighbour and next nearest neighbour FM interactions at the phase boundaries, as will be discussed in the following section. In addition, the present results may imply that there are metastable magnetic moment orientations out of the mirror plane as well as stable magnetic moment orientations within the mirror plane for the $Au_{64}Ga_{22}Tb_{14}$ 1/1 AC.

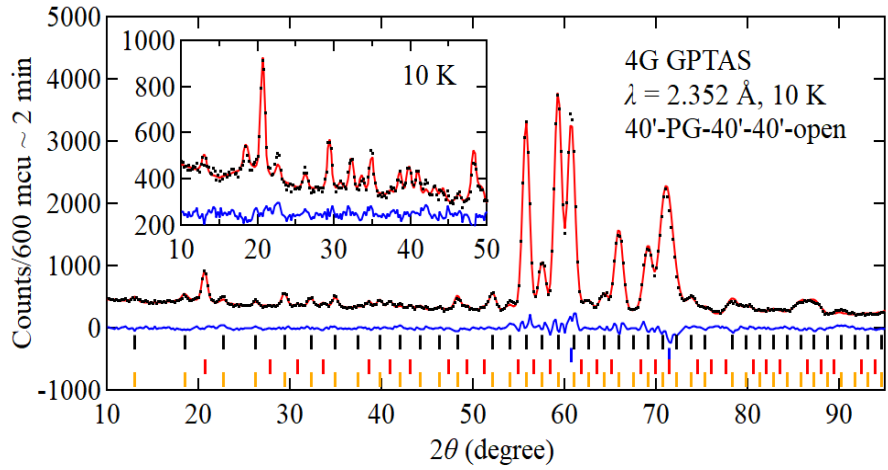

FIG. 12. Rietveld refinement of $Au_{64}Ga_{22}Tb_{14}$ 1/1 AC (10 K). Observed intensities, calculated intensities, and their difference are represented by black dots, red and blue curves, respectively. The position of the nuclear reflections, Aluminum (from sample holder), magnetic reflections from the AFM phase, and those from the FM phase are indicated by black, blue, red, and orange solid lines, respectively. Inset shows the enlarged view below 50 degrees.

*3.2.3. Developing magnetic phase diagram*

Based on the above observations, a magnetic phase diagram of the non-Heisenberg Tsai-type 1/1 Au-Ga-Tb ACs within a wide $e/a$ range of 1.6 – 2.2 is developed, as shown in Fig. 13a, wherein the red and light grey markers correspond to transition temperatures ($T_N$ for AFM, $T_C$ for FM or $T_f$ for spin freezing) and $\theta_w$, respectively. Notice that the magnetic susceptibility and inverse magnetic susceptibility of the samples from which the datapoints in Fig. 13a have been extracted are provided separately in Figs. S4 and S5 in the Supplementary Information. Clearly, at $e/a < 1.72$, an AFM order with whirling spin configuration along a crystallographic [111] axis is stabilized. The magnetic structure corresponding to AFM order can be represented by the magnetic space group $I_p m'\bar{3}'$ (#204.5, OG setting). Such spin arrangement on the icosahedral clusters can be best illustrated along the 3-fold axis, as depicted in Fig. 13b. The spins are almost tangential to the cluster surface lying on the mirror plane with the angle between their position vector from the origin and their direction of approximately 86 degrees [18,24]. At $e/a > 1.72$, a FM order occurs, where the whirling arrangement of spins along the 3-fold axis is likely with the magnetic space group $R\bar{3}$ (see Fig. 13b). Here, the notation "whirling" has been given to both ordered states because the spins on the icosahedron show a whirling pattern when viewed along the three-fold axis. Such whirling orders are due to the existence of a threefold axis in both magnetic space groups, because of which spontaneous magnetization is only allowed along the threefold direction for the FM phase.

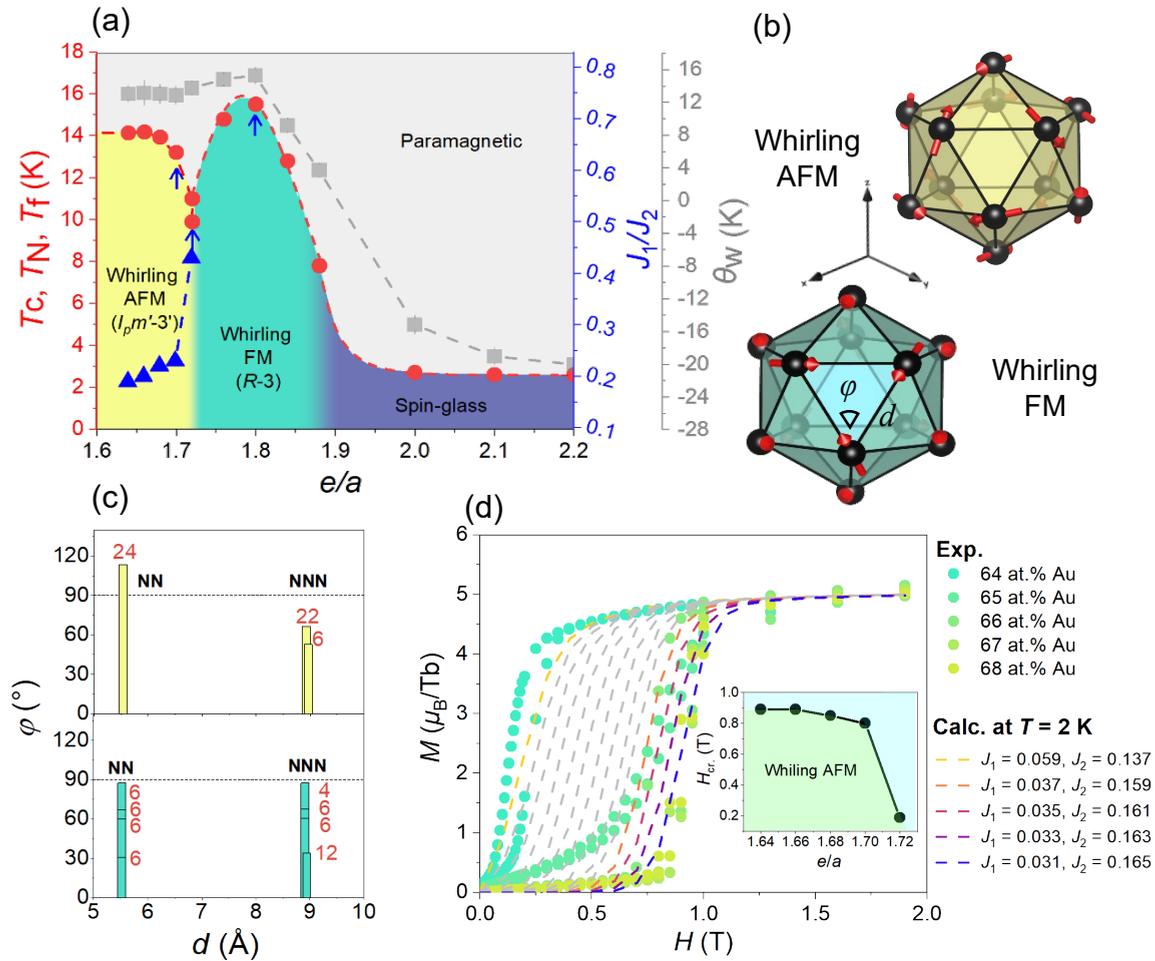

FIG. 13. (a) A magnetic phase diagram of the Au-Ga-Tb 1/1 ACs showing $e/a$ dependence of $T_C$, $T_N$ or $T_f$ (red markers), $\theta_w$ (grey markers) and $J_1/J_2$ (blue markers). The yellow, cyan, and dark blue background colors represent whirling AFM, whirling FM, and spin-glass regimes, respectively. The corresponding magnetic structures of the whirling AFM and FM orders are shown in (b). Blue arrows in Fig. 13a represent the samples studied by the PND experiment. (c) represent angular distributions (in degrees) of spins ($\varphi$) in the whirling AFM and FM phases, respectively, versus the distance between the RE elements symbolled as $d$. (d) Series of experimental and calculated field ($H$) dependence magnetizations of AFM samples with various $e/a$. The inset of (d) shows the $e/a$ dependence of critical magnetic field ($H_{cr.}$) above which the AFM state is destabilized.

At $e/a = 1.72$, i.e., exactly at the border of the FM/AFM states, a whirling AFM order isostructural to that of $Au_{65}Ga_{21}Tb_{14}$ 1/1 AC [24] is found to be a ground state at low temperatures, while in the same compound within a temperature range of 9.9 K – 11.2 K, both AFM and FM orders observed in the samples with $e/a = 1.70$ and $e/a = 1.80$, respectively, are recognized. Interestingly, the transition temperature associated with [paramagnetic → FM or AFM], as represented by red dashed line in Fig. 13a, shows a significant drop at $e/a = 1.72$ indicating a competing nature between the two ordered states at the border. At higher $e/a$ values than ~ 1.87, spin-glass regime appears wherein both $T_f$ and $\theta_w$ values drop significantly, indicating much stronger competition in AFM interactions or increased chemical disorder in this regime.

Figures 13c displays the angular distributions (in degrees) of spins ($\varphi$) in the whirling AFM and FM phases as a function of the distance between the RE elements. The calculations were performed within a range of interatomic distances from 5 to 10 Å, which encompasses nearest neighbor (NN) and next-nearest neighbor (NNN) interactions in a 12-spin icosahedron cluster. To avoid duplications, each angle between spins A and B was counted only once. The numbers beside each column indicate the frequency of occurrence within the distance range, including 52 distinct angles between spins, 24 corresponding to NN and the rest to NNN. The angle distributions provide valuable insights, revealing that the NNN interaction ($J_2$) is predominantly ferromagnetic ($\varphi < 90°$) in both whirling AFM and FM phases. This observation aligns with the positive values of $\theta_w$ observed in the FM and AFM regimes in Fig. 13a. Conversely, the NN interaction ($J_1$) in the whirling AFM and FM phases is predominantly antiferromagnetic ($\varphi > 90°$) and ferromagnetic ($\varphi < 90°$), respectively. Hence, it can be inferred that the magnitude of $J_1$ likely plays a crucial role in determining the phase selection between the whirling FM and AFM phases.

One effective approach to further support the discussion about the relationship between $J_1/J_2$, $e/a$ and the magnetic ground state is to analyze the magnetization curves of the AFM 1/1 ACs with varying $e/a$, focusing on the critical magnetic field ($H_{cr.}$) that corresponds to meta-magnetic-like anomalies commonly observed in AFM phases. The magnetization curves of the $Au_{68-x}Ga_{18+x}Tb_{14}$ ($x = 0-4$) 1/1 ACs falling within the AFM region of the magnetic phase diagram are presented in Fig. 13d, exhibiting their response to applied magnetic fields. The inset graph reveals a clear exponential decrease in $H_{cr.}$ as $e/a$ increases from 1.64 to 1.72.

The experimental magnetization curves in Fig. 13d can be well reproduced by performing calculations on a single icosahedron cluster composed of 12 Ising moments and Heisenberg interactions, as described in previous studies [18,24,25]. The calculations utilize the following Hamiltonian:

Table I. Nominal compositions, electron per atom (e/a), Néel temperature ($T_N$), critical magnetic field ($H_{cri}$) and estimated $J_1/J_2$ values for the Au-Ga-Tb 1/1 ACs

| Composition | e/a | $T_N$ (K) | $H_{cri.}$ (T) | $J_1/J_2$ |
|---|---|---|---|---|
| $Au_{68}Ga_{18}Tb_{14}$ | 1.64 | 14.13 | 0.89 | 0.19 |
| $Au_{67}Ga_{19}Tb_{14}$ | 1.66 | 14.18 | 0.89 | 0.20 |
| $Au_{66}Ga_{20}Tb_{14}$ | 1.68 | 13.93 | 0.85 | 0.22 |
| $Au_{65}Ga_{21}Tb_{14}$ | 1.70 | 13.20 | 0.80 | 0.23 |
| $Au_{64}Ga_{22}Tb_{14}$ | 1.72 | 9.92 | 0.19 | 0.43 |

$$H = -J_1 \sum_{NN} S_i \cdot S_j - J_2 \sum_{NNN} S_i \cdot S_j + g\mu_B H \sum_i S_i . \qquad (2)$$

Here, $J_1$ and $J_2$ represent the NN and NNN magnetic interactions, respectively. In Fig. 13d, a series of calculated magnetization curves, represented by dashed lines, are plotted alongside the experimental data for comparison. The $J_1/J_2$ ratio ranges from 0.19 to 0.43 in the calculations, with the former and latter corresponding to e/a = 1.64 and 1.72, respectively. The estimated $J_1$ and $J_2$ values presented in Fig. 13d not only accurately replicate the experimental $H_{cr.}$ values, but also yield a Weiss temperature of 13.79 K using $\theta_w = 5J(J + 1)(J1 + J2)/3$ ($J = 6$), which is in perfect accordance with the experimental $\theta_w$ values of the AFM samples (e/a ≤ 1.72). Table I lists the estimated $J_1/J_2$ ratios and the corresponding $H_{cr.}$ values for each e/a parameter. The results indicate that the $J_1/J_2$ ratio rapidly increases at the phase boundary of the AFM and FM phases and the whirling AFM ground state becomes unstable when $J_1$ exceeds a threshold value of $J_1 = J_2/2$ (see [18,24,25] for details about the threshold value). Consequently, a rapid increase of $J_1$ competing with $J_2$ can induce the dip in the transition temperature observed in Fig. 13a at e/a = 1.72. Indeed, the two-phase coexistence was observed at this e/a (see Fig. 10) indicating that the stability of the AFM and FM phases is very close. Overall, the magnetic phase diagram of the Au-Ga-Tb 1/1 ACs, developed in this study, offers valuable insights into the intricate interplay between magnetic interactions in non-Heisenberg Tsai-type 1/1 ACs.

## 4. Conclusion

The first comprehensive magnetic phase diagram of the non-Heisenberg Tsai-type AC is developed as a function of electron-per-atom ($e/a$) ratio, based on the bulk magnetization and powder neutron diffraction (PND) results. The PND measurements clearly demonstrated the occurrence of a noncoplanar whirling AFM order as the ground state at $e/a = 1.72$ and a presence of a noncoplanar whirling FM order at $e/a = 1.80$, both with magnetic moments tangential to the Tb icosahedron. Using numerical calculations on a non-Heisenberg single icosahedron, the FM/AFM phase selection rule was successfully explained in terms of the nearest neighbour ($J_1$) and next nearest neighbour ($J_2$) interactions. These findings provide an important foundation for understanding the intriguing magnetic orders of not only non-Heisenberg ACs but also non-Heisenberg $i$QCs that are yet to be discovered.

**Author contributions**

F.L. and K.N. designed, conducted the experiments, collected, and analysed the data and wrote the manuscript. S.S., K.I., H.W., and K. K assisted in performing part of the experiments. A.I. helped in conceptualization and material synthesis. T.F. assisted in performing specific heat measurements. T.S. and R.T. supervised the experiments. R.T. funded the experiments, wrote, review and edit the manuscript.

**Declaration of competing interests**

The authors declare that they have no known competing financial interests or personal relationships that could have appeared to influence the work reported in this paper.

**Acknowledgment**

This work was supported by Japan Society for the Promotion of Science through Grants-in-Aid for Scientific Research (Grants No. JP19H05817, JP19H05818, JP19H05819, JP21H01044,

JP22H00101, JP22H04582) and Japan Science and Technology agency, CREST, Japan, through a grant No. JPMJCR22O3. The experiments at JRR-3 were supported by the General User Program for Neutron Scattering Experiments (No. 22511), Institute for Solid State Physics, University of Tokyo and Sumitomo Foundation. The authors acknowledge Dr. Kazuyasu Tokiwa from Tokyo University of Science for his assistance in performing magnetic susceptibility measurements on some of the studied samples.